\documentclass[aps,prl,twocolumn,superscriptaddress,showpacs]{revtex4-1}
\usepackage{graphicx}
\usepackage{bm}
\begin{document}

%\preprint{}

\title{Spin Ice: Magnetic Excitations without Monopole Signatures using Muon Spin Rotation}
  \author{S.~R.~Dunsiger}
 \email{sarah.dunsiger@frm2.tum.de}  
  \affiliation{Physik Department, Technische Universit\"{a}t M\"{u}nchen, D-85748 Garching, Germany}
    \author{A. A. Aczel}
     \affiliation{Department of Physics and Astronomy, McMaster University, Hamilton, Ontario L8S 4M1, Canada}
     \author{C.~Arguello}
     \affiliation{Department of Physics, Columbia University, New York, New York 10027, USA}    
  \author{H. Dabkowska}
     \affiliation{Brockhouse Institute for Materials Research, McMaster University, Hamilton, Ontario L8S 4M1, Canada}
 \author{A.~Dabkowski}
     \affiliation{Brockhouse Institute for Materials Research, McMaster University, Hamilton, Ontario L8S 4M1, Canada}
 \author{M.-H.~Du}
 \affiliation{Materials Science and Technology Division, Oak Ridge National Laboratory, Oak Ridge, Tennessee 37831-6114, USA}
     \author{T. Goko}
     \affiliation{Department of Physics, Columbia University, New York, New York 10027, USA}
     \affiliation{TRIUMF, 4004 Wesbrook Mall, Vancouver, British Columbia, V6T 2A3, Canada}
     \author{B.~Javanparast}
     \author{T.~Lin}
 \affiliation{Department of Physics and Astronomy, University of Waterloo, Waterloo, Ontario, N2L 3G1, Canada}
 \author{F.~L.~Ning}
      \affiliation{Department of Physics, Columbia University, New York, New York 10027, USA}
      \affiliation{Department of Physics, Zhejiang University, Hangzhou 310027, China}
   \author{H.~M.~L.~Noad}
     \affiliation{Department of Physics and Astronomy, McMaster University, Hamilton, Ontario L8S 4M1, Canada}
 \author{D.~J.~Singh}
 \affiliation{Materials Science and Technology Division, Oak Ridge National Laboratory, Oak Ridge, Tennessee 37831-6114, USA}
      \author{T.~J.~Williams}
     \affiliation{Department of Physics and Astronomy, McMaster University, Hamilton, Ontario L8S 4M1, Canada}  
    \author{Y. J. Uemura}
    \email{tomo@lorentz.phys.columbia.edu}
    \affiliation{Department of Physics, Columbia University, New York, New York 10027, USA} 
\author{M.~J.~P.~Gingras}
     \email{gingras@gandalf.uwaterloo.ca}
 \affiliation{Department of Physics and Astronomy, University of Waterloo, Waterloo, Ontario, N2L 3G1, Canada}
     \affiliation{Canadian Institute for Advanced Research, Toronto, Ontario M5G 1Z8, Canada}  
         \author{G.~M.~Luke}
     \email{luke@mcmaster.ca}
     \affiliation{Department of Physics and Astronomy, McMaster University, Hamilton, Ontario L8S 4M1, Canada}
     \affiliation{Canadian Institute for Advanced Research, Toronto, Ontario M5G 1Z8, Canada}    
      \date{\today}
   \begin{abstract}
Theory predicts the low temperature magnetic excitations in spin ices consist 
of deconfined magnetic charges, or monopoles. A recent transverse-field (TF) muon spin 
rotation ($\mu$SR) experiment  [S T Bramwell {\it et al.}, Nature {\bf 461}, 956 (2009)] reports results claiming to be consistent with the 
temperature and magnetic field 
dependence anticipated for monopole nucleation $-$ the so-called second Wien effect.  
We demonstrate via a new series of $\mu$SR experiments in Dy$_2$Ti$_2$O$_7$ that such an effect is not observable in a TF $\mu$SR experiment. Rather, as found in many highly frustrated magnetic materials, we observe spin 
fluctuations which become temperature independent at low temperatures, behavior which dominates over any possible signature of thermally nucleated monopole excitations.
\end{abstract}
%\keywords{}
\maketitle
%%%%  New Introduction %%%%

Spin ices, such as Ho$_2$Ti$_2$O$_7$ and Dy$_2$Ti$_2$O$_7$,
 are topical highly frustrated magnetic systems which 
 exhibit a gamut of very interesting phenomena~\cite{Harris:1997p3030,Bramwell:2001p5742}. 
In (Ho/Dy)$_2$Ti$_2$O$_7$, the Ho$^{3+}$ and Dy$^{3+} $ magnetic ions reside on the vertices of
a pyrochlore lattice of corner-sharing tetrahedra. 
A large single ion anisotropy forces the moment  to point strictly along local $\langle111\rangle$ crystalline axes,
 along the line which connects the centers of the two adjoining tetrahedra and their common vertex, 
making the moments classical ``local'' Ising spins.
Since  Ho$^{3+}$ and Dy$^{3+}$  carry a large magnetic moment of $\sim 10\ \mu_{\rm B}$,
the  dipolar interaction in these systems is $\sim 1$~K at nearest neighbor distance
and of similar magnitude as  the Curie-Weiss temperature $\theta_{\rm CW}$ \cite{Bramwell:2001p5742}.
 The frustration in spin ices  stems from the  $1/r^3$ 
 long-range nature of the magnetic dipolar interaction and its consequential ``self-screening''
 ($r$ is the distance between ions)~\cite{Gingras:2001p5949,Isakov:2005p6131,Castelnovo:2008p5463}.
As a result, spin ices are frustrated ferromagnets 
with low-energy states characterized by  
two spins ``pointing in'' and two spins ``pointing out'' on each tetrahedron --
the  2-in/2-out  rule which defines minimum energy spin configurations. 
These map onto the allowed 
proton configurations in water ice which
obey the Bernal-Fowler ice-rules~\cite{Bramwell:2001p5742}, hence the name spin ice.
 
The   dipolar spin ice model~\cite{denHertog:2000p5822}
 and its refinement \cite{Yavorskii:2008p2967} yield an accurate microscopic quantitative description of the 
 equilibrium thermodynamic
properties of spin ices, both in zero and nonzero magnetic 
field.  In contrast, the problem of the dynamical response of the
moments  in spin ices remains much less studied and understood.
An exciting recent development in that direction 
is the realization that the ``2-in/2-out'' spin configurations
may be described via a divergence-free coarse-grained 
 magnetization density field \cite{Castelnovo:2008p5463,Ryzhkin:2005wn}.
A thermal fluctuation causing the flip of an Ising spin 
from an ``in'' to an ``out'' direction amounts to the creation
of a nearest-neighbor pair of magnetization source and sink on
the two adjoining tetrahedra  or, in other words,  to the 
nucleation of  ``magnetic monopoles'' out of  the  spin-ice-rule obeying vacuum \cite{Castelnovo:2008p5463}.
Particularly interesting is the observation that monopoles in dipolar spin ice interact via
an emerging Coulomb potential which decays inversely proportional to the distance which
separates them and are therefore deconfined~\cite{Castelnovo:2008p5463}.
A recent numerical study \cite{Jaubert:2009p2960} provides
evidence that the temperature dependence of the relaxation time determined
in ac magnetic susceptibility measurements~\cite{Snyder:2004p3003}
can be rationalized in terms of thermally activated monopoles, at least above ~1 K.
The wave vector dependence of the neutron scattering intensity suggests power law spin correlations,
which are a prerequisite for monopoles with effective Coulomb interactions \cite{Fennell:2009p2957}.
Yet, perhaps the reported direct evidence for the presence of monopoles in spin ice and a
determination of their effective charge is the most intriguing recent result \cite{Bramwell:2009p2963}.

In weak electrolytes, including water ice, characterized by a small dissociation rate
constant $K$, the so-called second Wien effect describes the nonlinear 
increase of $K$ under an applied electric field.
In a recent paper \cite{Bramwell:2009p2963},  Bramwell and co-workers 
have drawn further on the analogy between magnetic moments in spin ice 
and protons in water ice~\cite{Bramwell:2001p5742}.
Using Onsager's accurate theory of the Wien effect \cite{Onsager:1934p6571},  
Bramwell {\it et al.} put forward an elegant  
model to describe the dependence of the monopole nucleation rate, $\kappa(T,H)$ in spin ice
on temperature $T$ and external applied magnetic field $H$. 
They proposed that a measurement of
$\kappa(T,H)$ could yield  the monopole charge $Q$.
To that effect, the authors of Ref.~\cite{Bramwell:2009p2963} used
$\mu$SR in a transverse-field (TF) geometry to determine $\kappa(T,H)$ and extract a value  $Q_{{\rm exp}} \sim 5\mu_{{\rm B}}/$ \AA, close to the
value  $Q_{{\rm theo}} \sim 4.6\mu_{{\rm B}}/$ \AA\ anticipated by theory~\cite{Castelnovo:2008p5463}.

In this Letter, we discuss how the weak TF $\mu$SR experiment
of Ref.~\cite{Bramwell:2009p2963}, as a means to observe the second Wien effect in spin ice,
was flawed in its conceptual design and execution and 
incorrect in its theoretical interpretation of the  muon spin depolarization rate.
Monte Carlo calculations show that the internal magnetic field
at the expected muon locations in Dy$_2$Ti$_2$O$_7$ spin ice material is ``large''  
($\sim0.3$~T) and has a broad distribution, preventing the observation of TF muon precession.  We present evidence
 that the coherent muon precession in weak ($H\sim1$~mT) TF  seen  in Ref.~\cite{Bramwell:2009p2963} 
 originated rather from the sample holder and other parts 
of the sample environment. In contrast, our zero-field $\mu$SR results
exhibit low-temperature muon spin relaxation which is temperature independent from
4 K down to 20 mK. 

Positive muons provide a pointlike real space magnetic probe averaging over the Brillouin zone, in contrast
with magnetization measurements, which measure only the Q=0 response.  In a $\mu$SR experiment, essentially 100\% spin--polarized positive
 muons are implanted in a material and precess in the local magnetic field ${\mathbf B}({\mathbf r})$.  The muons subsequently decay (with lifetime $\tau_\mu=2.2\;\mu s$) into a positron (emitted preferentially in the direction of the muon spin at the time of decay) and two undetected neutrinos.  An asymmetry signal, obtained from the decay histograms of opposing positron detectors, represents the projection of the muon spin polarization function onto the axis defined by the detectors.  Since the muons are created fully spin polarized (via the parity-violating weak decay of their parent pions), $\mu$SR experiments may be performed in zero (ZF), longitudinal (LF) or transverse (TF) magnetic field.  As in nuclear magnetic 
resonance (NMR),
 the depolarization results from both dynamic ($T_1$) and static ($T_2$) processes.  Further details
 describing $\mu$SR are found elsewhere~\cite{scotland_book}.  

TF-$\mu$SR -- In a TF experiment, an external magnetic field is applied perpendicular to the initial muon spin 
polarization direction.  
For arbitrary electronic spin fluctuation rates, an approximate analytic form for the high TF-$\mu$SR
polarization function is given by a relaxation envelope multiplied by a cosine precession 
signal~\cite{scotland_book,hayano79},
\begin{equation}
{P}_\mu(t) =  \exp\left[-(\Delta^2/\nu^2) \left(e^{-\nu t}-1+\nu t\right)\right]\cos(\omega t).
\end{equation}
In Eq. (1), $\Delta=\gamma_\mu B$ is the muon gyromagnetic ratio times the rms instantaneous internal magnetic field at the muon site $B$ and $\nu$ is the field fluctuation rate.  In the fast fluctuation
($\nu\gg\Delta$)  regime, the envelope becomes $\exp(-\Delta^2t/\nu)$ whereas,  in the slow fluctuation 
($\nu\ll\Delta$) regime, the  envelope
reduces to $\exp(-\Delta^2t^2/2)$, which is {\em  independent} of $\nu$.  Hence, the 
relaxation in a transverse field (TF) never takes an exponential form where the 
relaxation rate $1/T_1\propto \nu$, as assumed by 
Bramwell {\em et al.} \cite{Bramwell:2009p2963}. As discussed further below, such
 behavior is only 
applicable to ZF and LF measurements.
 As the fluctuation rate of the electronic moments decreases and a significant spectral 
 density develops near zero frequency, the local field at the muon site becomes the vector sum of the applied and internal 
 fields.
 As we show, this net field in dipolar spin ice has a much larger rms value than the applied field ${\bm H}$.
Therefore the muon polarization function is 
 not given by a cosine with a frequency corresponding to the applied field but, instead, is rapidly damped to zero.  

From the discussion above, a crucial issue is whether the internal field distribution $P({B}({\bm r}))$ in a spin ice has
 significant weight below the applied external field of $H\sim 1$ mT.
 As a first step to address this question, we use a loop algorithm \cite{Melko:2004hk} in Monte Carlo (MC) simulations 
 of a realistic microscopic model of Dy$_2$Ti$_2$O$_7$ \cite{Yavorskii:2008p2967} to calculate
 $P({B}({\bm r}))$ at the most probable muon locations, as determined by density functional
 theory (DFT) calculations~\cite{supp_mat}. 
 We find, confirming naive expectations, that $P({B}({\bm r}))$ is heavily populated
 for fields of several hundreds of millitesla. Such values
 are consistent with the estimate of $\sim0.5$~T obtained by Lago {\it et al.} from LF- $\mu$SR 
 decoupling measurements in Dy$_2$Ti$_2$O$_7$ \cite{Lago:2007p2979}.  In fact, for all four lowest energy potential muon stopping sites, we find vanishing $P({B}({\bm r}))$ at $\vert  ({B}({\bm r})) \vert \rightarrow 0$.  

\begin{figure}[t]
\includegraphics[angle=0,width=\columnwidth]{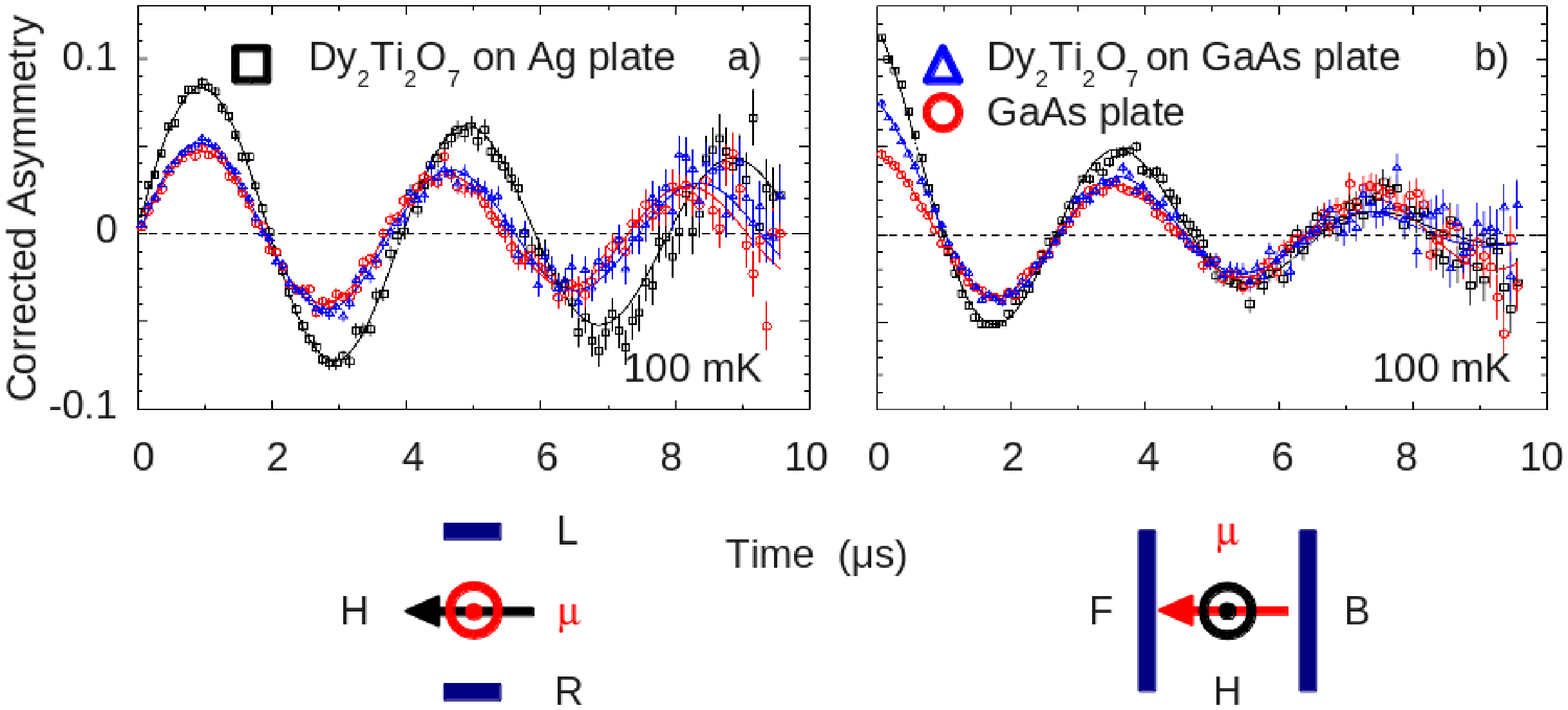} 
\caption{\label{tf_data}(Color online) TF $\mu$SR spectra measured at $T=100$~mK in Dy$_2$Ti$_2$O$_7$ with two different counter geometries.  In
(a), the applied field $H=2$ mT lies in the plane of the platelike coaligned mosaic of single crystals.  In (b), $H$ is perpendicular to the plates.  
In both cases $H$ is along the crystallographic [100] direction. The positron detectors are indicated as back ($B$), forward ($F$), left ($L$) and right ($R$).}
\end{figure}

Figures~\ref{tf_data}(a) and \ref{tf_data}(b) show TF-$\mu$SR spectra measured in single crystals of Dy$_2$Ti$_2$O$_7$  
mounted using GE varnish on an intrinsic GaAs plate 
(blue triangles), as well  as results on a blank GaAs plate without the Dy$_2$Ti$_2$O$_7$ sample (red circles) with an 
external field of 2~mT in both cases.  The crystals were grown using
floating zone image furnace methods, with the speed of 4 mm/h.  
GaAs was chosen as it exhibits no precession at the muon Larmor frequency
 since all muons form muonium, a hydrogenlike muon-electron bound pair~\cite{Kiefl:1985hz}.
As a result, the observed red precession signal in Figs.~\ref{tf_data}(a) and \ref{tf_data}(b) is a purely instrumental 
background from muons which land elsewhere in the dilution refrigerator or in  the silver (Ag) sample holder.
  The signal with the Dy$_2$Ti$_2$O$_7$ sample on the GaAs plate (blue) is essentially 
identical to the background contribution in Fig.~\ref{tf_data}(a) at long ($t\gtrsim1\;\mu{\rm s}$) times. 
 This demonstrates that there is no long-lived 
muon precession signal originating from the specimen, consistent with the
aforementioned expectation based on DFT-MC calculations~\cite{supp_mat}.  The small applied $H=2$~mT  is hence negligible compared to $B({\bm r})$.  The  small difference in Fig.~\ref{tf_data}(b) at early 
times ($t\lesssim1\;\mu$s) comes from the longitudinal relaxation of the $\mu$SR signal in the Dy$_2$Ti$_2$O$_7$ sample, 
visible here because the initial muon spin polarization is parallel to the positron detector axis.  This difference is absent 
in Fig.~\ref{tf_data}(a) where, due to the chosen  experimental geometry, the initial muon polarization is perpendicular to the detector axis.
TF-$\mu$SR spectra measured with the Dy$_2$Ti$_2$O$_7$ crystals mounted on a 
high purity Ag plate for the same two experimental geometries are shown in black.  Using a metal (e.g. silver) backing plate ensures the
sample is in good thermal contact.  Additionally, silver (taken alone)
produces an essentially undamped TF-$\mu$SR precession signal. The increased signal amplitude relative 
to the signal using a GaAs backing reflects the contribution of muons landing in parts of the Ag backing not covered by the sample.  Such TF-$\mu$SR spectra measured using Ag are essentially identical to those reported in 
Ref.~\cite{Bramwell:2009p2963}.  Note that in both instances, to account for history dependent effects,
care was taken to cool the samples in zero field before applying 2 mT at
base temperature and taking data while warming.
We note that the evolution of the TF depolarization rate tracks the magnetization~\cite{Snyder:2004p3003}. 
We speculate that a stray field, proportional to the dc magnetization, is generated within the Ag plate in the 
areas between the crystallites of Dy$_2$Ti$_2$O$_7$.  Muons landing in such a region undergo slow relaxation due 
to  the inhomogeneous stray field, proportional to the applied field, consistent with observations
of Bramwell {\em et al.}~\cite{Bramwell:2009p2963}.

\begin{figure}[t]
\includegraphics[angle=0,width=\columnwidth]{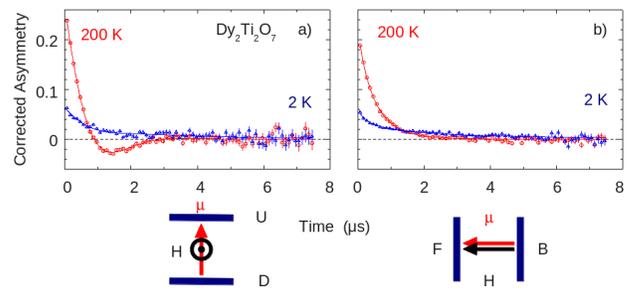} 
\caption{\label{low_back}(Color online) Background free (a) TF=2 mT  and 
(b) ZF $\mu$SR spectra measured 
in Dy$_2$Ti$_2$O$_7$.  Note the overdamped precession signal in (a).  The positron detectors are indicated as back ($B$), forward ($F$), up ($U$) and down ($D$).}
\end{figure}

The above experiments, performed in a dilution refrigerator,
require a cold finger sample holder to provide thermal contact to the sample.  In contrast, by using a $^4$He 
gas-flow cryostat operating at temperatures above $T=2$~K, one can perform $\mu$SR without complications due to a
background signal  
by suspending the specimen on thin tape.  The TF-$\mu$SR signal observed in such a ``background free'' apparatus 
is shown in Fig.~\ref{low_back}(a) in $H=2$~mT.  The absence of a precession signal at $T=2$~K 
confirms that any long lived precession signal seen at $T\lesssim2$~K
does not originate from Dy$_2$Ti$_2$O$_7$.  Rather, it is a background signal from the sample holder or cryostat.  
This constitutes the first of our two main results.

\begin{figure}[t]
\includegraphics[angle=90,width=7cm]{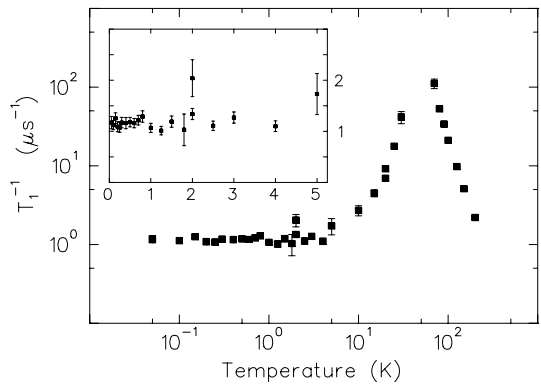} 
\caption{\label{t_1} Muon spin relaxation rate in Dy$_2$Ti$_2$O$_7$.  Individual $\mu $SR spectra were analyzed using
a phenomenological stretched exponential form $P_Z(t) \propto \exp(-(t/T_1)^{\beta })$ commonly used to model glassy systems.  
The low temperature behavior of T$_1^{-1}$ is shown on an expanded linear scale in the inset.  
The exponent $\beta $ drops monotonically from $\sim 0.75$ at 150 K to 0.4 below 5 K.}
\end{figure}

ZF-$\mu$SR -- A more effective method for studying spin dynamics in magnetic systems is to measure the 
spin polarization in the ZF/LF geometry~\cite{scotland_book}.  At high temperatures, in the fast fluctuation regime, the spin lattice 
relaxation rate is $1/T_1=2\Delta^2/\nu$ in zero applied field~\cite{scotland_book}.   As shown in Fig.~\ref{t_1}, $1/T_1$  
increases as temperature decreases.  This is due to the combined effect of changes in the size of $\Delta$ as the 
Dy$^{3+}$ excited crystal electric field levels are depopulated and the slowing down of the Dy$^{3+}$ fluctuation 
rate  $\nu$.  The relaxation rate peaks at $T \approx 50$~K and drops below, entering the slow fluctuation regime, consistent with earlier studies on powder 
 samples \cite{Lago:2007p2979}.  Below $T=50$~K  the muon spin polarization exhibits a two component form, indicating that the local magnetic environment 
 consists of a large quasistatic field [responsible for the rapid loss of polarization seen in Fig.~\ref{low_back}(b) at 2~K] coexisting with a fluctuating field component 
 (which gives the damping of the remaining polarization).  The amplitude of the slowly relaxing component at $T \lesssim 50$ K
 is 1/3 of the initial polarization, as expected for a cubic material where 1/3 of the muon polarization is on average parallel to ${\mathbf B} ({\mathbf r})$~\cite{scotland_book}.
Analogous decreases in Dy$^{3+}$ fluctuation rate over 4 orders of magnitude between 300 and 8 K have been observed using 
zero field $^{47}$Ti NQR~\cite{kitagawa} and nuclear forward scattering~\cite{sutter}.

We observe a substantial $1/T_1\sim1\;\mu{\rm s}^{-1}$ relaxation rate at $T \lesssim 5$ K, more than an
order of magnitude above our detection limit ($\sim10^{-2}\;\mu{\rm s^{-1}}$). This observation contrasts dramatically with the activated behavior anticipated for magnetic 
 monopoles \cite{Jaubert:2009p2960} and is startling given the highly Ising nature of the Dy$^{3+}$ spins, where large energy barriers against single spin flip processes separate quasidegenerate ice rules states~\cite{Bramwell:2001p5742}.  The origin of this temperature 
 independent relaxation is as yet unclear.  We note that so-called persistent spin dynamics have been observed in a wide range of geometrically frustrated materials~\cite{uemura_scgo,McClarty:2011km}.
Oddly, the characteristic rare earth spin fluctuation rates extracted using various techniques differ dramatically 
in Dy$_2$Ti$_2$O$_7$, as do the values when comparing the two isostructural 
compounds A$_2$Ti$_2$O$_7$ (A=Dy,Ho)~\cite{Snyder:2001p5855,Matsuhira:2001cp,Snyder:2004p3003,Matsuhira:2000p3027,Clancy:2009p2964,Ehlers:2004iu}.
The strong hyperfine interaction 
 between the electronic and nuclear spin species should also not be neglected, particularly in the latter compound, as 
 highlighted by the pronounced Schottky anomaly observed arising from nuclear contributions to the magnetic 
 specific heat~\cite{Bramwell:2001p3025}. 

Dy-based compounds form a variety of model Ising
systems~\cite{Wolf:2000dg}. However, many of them exhibit unexpected
dynamic spin fluctuations at low temperatures: (1)
the single molecular magnetic system [DyPc$_2$]$^0$, characterized
by a doubly degenerate ground state and large
magnetic anisotropy, exhibits a
tunnelling regime~\cite{Branzoli:2010ee}, suggesting such temperature independent behavior is a more 
pervasive phenomenon~\cite{Vernier:2003he}; (2) the geometrically 
frustrated Ising antiferromagnet
Dysprosium Aluminium Garnet, where marked
changes in the characteristic relaxation times over several
orders of magnitude have been reported as a function of applied
fields~\cite{Wolf:2000dg}; (3) even the archetypal dilute Ising
dipolar ferromagnet Dysprosium Ethyl Sulphate, which exhibits unexpectedly high relaxation rates within the
ordered state~\cite{cooke68}.

Despite a seemingly compelling argument for
spin dynamics caused by monopoles \cite{Castelnovo:2008p5463,Jaubert:2009p2960}, 
additional spin relaxation processes dominate the behavior observable using  ZF-$\mu$SR in Dy$_2$Ti$_2$O$_7$ spin ice.  
In Heisenberg spin systems, 
geometrical frustration may lead to spin liquid behavior~\cite{Balents:2010p3694} where spin fluctuations persist to absolute zero. 
Understanding the low-temperature dynamics in the Ising Dy$_2$Ti$_2$O$_7$ system will require the construction of an 
effective low-energy Hamiltonian containing non-Ising terms~\cite{Molavian2007}.
The present work clarifies the nature of static and dynamic
contributions to the internal magnetic field as probed using $\mu$SR and
describes evidence of unusual spin excitations in a model Ising system.
It poses the challenge to comprehensively understand the microscopic mechanism(s) causing the temperature independent muon spin relaxation
within the broader context of geometrically frustrated systems~\cite{McClarty:2011km}.

{\bf Acknowledgement:\/}  We acknowledge technical support at TRIUMF from D. Arseneau and B. Hitti. This work has been supported 
by NSERC and CIFAR at McMaster University and the
University of Waterloo, the Canada Research Chair
(MJPG, Tier 1) and the U.S. NSF under MWN DMR-
0806846; PIRE OISE-0968226 and DMR-1105961 programs
at Columbia University. Work at ORNL was
supported by DOE, BES Materials Science and
Engineering Division. F. L. N. is supported by National
Basic Research Program of China (973 Program) under
Grant No. 2011CBA00103.
%\bibliographystyle{apsrev4-1}
%\bibliography{spin-ice-refs}

\begin{thebibliography}{30}
\bibitem{Kramers} Dy$^{3+}$ is an odd-electron ion and, thanks to Kramers' theorem,
the lowering of the local symmetry of the crystal-field will not lift the degeneracy 
of the single-ion crystal field ground doublet.
\end{thebibliography}

%--------------------------------------------------------------------------------------------------

\section{Supplementary Material for:  Magnetic Excitations without Monopole Signatures using Muon Spin Rotation}

This document is a supplement to our main article, in which we provide some of 
the details of the density functional theory (DFT)
and Monte Carlo (MC) calculations which were carried out to determine the most probable
muon location sites in the Dy$_2$Ti$_2$O$_7$ spin ice compound, as well as the associated internal magnetic 
field distribution.

%\section {Density Functional Theory (DFT) and Monte Carlo Calculations}

\subsection{DFT calculations}

\begin{figure}[htbp]
\begin{center}
\includegraphics[width=9cm]{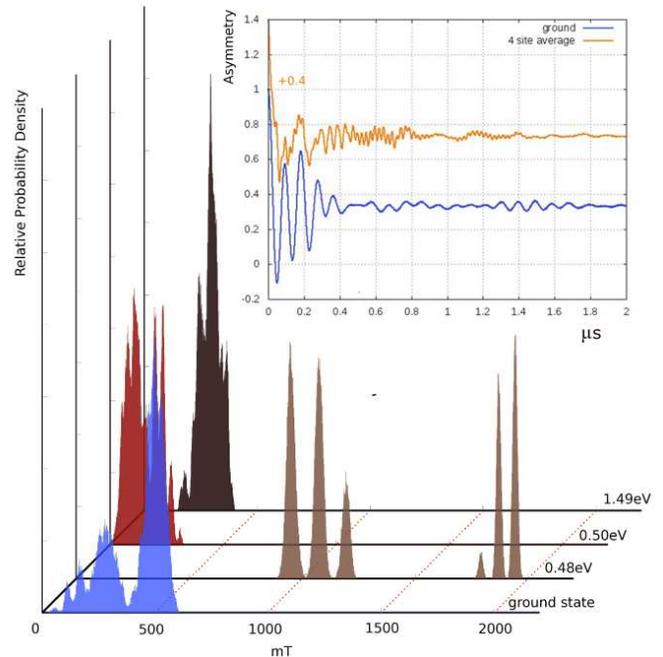}
\caption{(Color online).
Probability histograms of the internal magnetic fields at candidate muon sites 
are sampled by Monte Carlo (MC) simulations at $200$ mK.
 The insert shows the time-dependence of the muon spin polarization for both the muon location ground state
(curve labeled ground) and the average of the four lowest energy states. }
\label{supfig-1}
\end{center}
\end{figure}

To determine the quasistatic dipolar magnetic field which a positive muon experiences in the insulating Dy$_2$Ti$_2$O$_7$
spin ice compound, one must first determine the most probable local minimum energy locations a muon occupies after coming
to rest upon high (4.1 MeV kinetic) energy implantation in the sample.
The standard expectation 
is that a positive muon (a light version of proton) forms a (hydrogenlike) bond with an oxygen at a typical distance of 1 \AA. 
To address this question more quantitively, we determined those low-energy muon locations (ground state and a few
metastable states) using density functional theory (DFT) calculations using the VASP package \cite{VASP}.
In those calculations, one single positive muon was placed inside an 88-atom Dy$_2$Ti$_2$O$_7$ supercell with periodic boundary conditions,
which was neutralized by applying a uniform negative background charge.
The valence wavefunctions were expanded in a plane-wave basis with a cutoff energy of 
400 eV. A $2\times 2 \times 2$ grid was used for the $k$-point sampling of the Brillouin zone.
Perdew-Burke-Ernzerhof exchange-correlation functionals \cite{Perdew} were used 
and all atoms in the cell were relaxed to minimize the Feynman-Hellmann forces to below 0.01 ev/\AA.
The ground state as well as three other metastable higher energy states were identified, 
with their energy measured from the ground state labeled in Fig.~\ref{supfig-1}.
The determined ground state as well as the identified  metastable higher energy states of muon locations 
were taken as the most likely candidate for the muon locations in Dy$_2$Ti$_2$O$_7$.
\vspace{2mm}

\subsection{MC calculations}

The internal magnetic  fields at the muon sites due to the static Dy$^{3+}$ magnetic moments were next calculated.
To determine this field, we performed Monte Carlo (MC) simulations 
to sample the magnetic moment configurations at various  temperatures.
For these calculations, any
local structural distortions of the original crystal caused by the presence 
of the muon were ignored in the MC.
We considered the same extended dipolar spin ice model of Dy$_2$Ti$_2$O$_7$ as
in Ref.~[\onlinecite{yavorskii}]
  with exchange interactions up to third nearest neighbors and long-range 
magnetostatic dipole-dipole interactions.
We employed a system size of 432 spins with periodic boundary conditions. 
The dipole-dipole interactions among the Dy$^{3+}$ magnetic moments as well as the
local internal magnetic field at the muon sites determined by DFT were calculated 
using the Ewald summation technique. 
A loop update algorithm \cite{Melko} was employed to overcome the energy barriers of the single-spin-flip updates at temperatures 
below the Schottky peak in the specific heat of Dy$_2$Ti$_2$O$_7$, which occurs
 at a temperature $T_{\rm peak} \sim 1.1$ K. 
At least 1000 MC steps with loop updates \cite{Melko} per spin were used in equilibrating the system. The sampling
of the local dipolar magnetic field was done using 360,000 spin configurations.
We assume that the time scale for the spin dynamics is, deep in the spin ice regime at low temperature
($T\ll T_{\rm peak}$),
 much longer than the muon lifetime and that the muons implanted homogeneously
in the sample, one by one, experience static fields from these spin configurations with equal probability.
No muon-Dy$^{3+}$  transferred hyperfine field was taken into account in the present calculations.
The probability histograms of the internal magnetic field at 200 mK are plotted in the main panel
of Fig.~\ref{supfig-1}.
The multiple peaked structures are a combined consequence of (i) of the discreteness of the Ising spins and (ii) the fact 
that spin ices are not completely random as in an
Edwards-Anderson spin glass, but rather, obey the two-in/two-out ice
rules with the consequential dipolar correlations \cite{Henley}.

\vspace{2mm}

One observes that the probability density for
magnetic fields of order of hundreds of millitesla or higher is large at
the lowest energy  candidate muon sites determined by DFT.
As an independent check, we determined the internal magnetic field at the surface of a spherical
shell of radius of 1 \AA\  centered on oxygen locations and confirmed that such a 
magnitude of internal magnetic field is indeed typical.
 We also note that there is vanishing probability of
$B({\mathbf {r}}) \lesssim 10$ mT for all muon locations.
The inset of Fig.~\ref{supfig-1} shows the 
zero field (ZF) muon spin depolarization associated with the muon ground state (curve labeled
``ground'') as well as the average relaxation for the ground state and the
three lowest metastable states (curve labeled ``4 site average'')
with initial muon polarization along [100] (see main text).
The fast oscillations of the muon polarization
for short time ($t\lesssim 0.5$ $\mu$s) with its rapid decay and asymptotic
approach to a 1/3 value is again a reflection of 
the typical magnitude of the magnetic field a muon experiences, which is of order 100 mT.
Interestingly, the experimental $\mu$SR data do not find such
``front end'' oscillations, which should be observable given the time binning used in the
data acquisition and the instrumental (electronic) timing resolution.
This is possibly a result of the slight randomization of the local crystal field
Hamiltonian acting at the Dy$^{3+}$ ions neighboring the implanted muon, 
and hence the corresponding weakly randomized magnitude of the
Dy$^{3+}$ magnetic moments~\cite{Kramers}, as well as their direction
and which we have ignored. Another randomization which we have ignored is 
that of the transferred hyperfine field. 
Finally, and most notably, these DFT-MC calculations neglect the effect of 
Dy$^{3+}$ spin
fluctuations on the timescale of the muon lifetime $-$ perhaps the
most significant new result of our
experimental study and reported in the main part of the paper.  Indeed, as observed
experimentally, these have perhaps the most dramatic effect on the muon spin depolarization function.

While all these effects which go beyond
the approximations used here should ultimately be taken into account in future
calculations, the results of such improved calculations will not affect our main conclusion. 
Namely, the configuration of Dy$^{3+}$ magnetic moments in Dy$_2$Ti$_2$O$_7$ are
sufficiently random, despite the ice-rule, that the likelihood that
a finite percentage of muons land at sites with a local magnetic field
negligible compared to the applied external field of 2 mT is for all practical
purposes zero.
Ultimately, the experimental results reported in the main part of the paper
corroborate this assertion: no weak TF precession signal whose origin  is
the  Dy$_2$Ti$_2$O$_7$ sample can be positively identified, confirming the above
DFT-MC calculations that there are no density of (muon occupation) states
with local field of the order or less than the applied magnetic fields of
order of 2 mT.

%\addcontentsline{toc}{section}{References}

\end{document}